\documentstyle[12pt]{article}
\newcommand{\aslash}{\not\!\!A}
\newcommand{\dslash}{\not\!\partial}

\begin{document}
\begin{titlepage}
\begin{flushright}
hep-th/9409198 \\
IASSNS-HEP-94/70\\
UM-P-94/96\\
RCHEP-94/26\\
September, 1994
\end{flushright}

\vspace{0.9in}

\begin{center}
{\large\bf Chiral Gauge Theory in Four Dimensions}\\

\vspace{0.3in}

Tien D Kieu\\

\vspace{0.2in}

Institute for Advanced Study,\\
Princeton NJ 08540,\\
USA\\

\vspace{0.3in}

School of Physics~\footnote{Present address.},\\
University of Melbourne,\\
Parkville Vic 3052,\\
Australia

\vspace{0.2in}

\end{center}
\begin{abstract}
\noindent
A formulation of abelian and non-abelian chiral gauge theories is presented
together with arguments for the unitarity and renormalisability
in four dimensions.
\end{abstract}
\end{titlepage}
\noindent
Chiral gauge theory is so central to our understanding of the electroweak
interactions and yet is susceptible to the anomalous
violation of the very gauge symmetry defining the theory and, not unrelated,
a gauge-invariant, non-perturbative regulator is presently
nowhere to be found.  There is also the difficulty in generating
mass for the various participating fields, particularly since the existence
of the Higgs particles is not established.

On the one hand, anomaly cancelation can impose, from the outside,
stringent constraints for consistent and thus admissible theory.  On the other,
`anomalous' theory is interesting in itself and may not be inconsistent as
presumed.  There have been several attempts~\cite{1} trying to construct
meaningful `anomalous' theory.
This letter presents an explicit construction of four-dimensional
theory for both abelian and non-abelian gauge groups.  Arguments
are also provided for their consistency, unitarity and perturbative
renormalisability.  The construction is immediately  generalisable to any
even number of dimensions.
We will first present the motivation for the construction
of the path-integral action.  Once this explicit form is
obtained, we will take it as the starting point and investigate the
viability of the theory.

It was observed that in the holomorphic representation
the Berry's phase accompanying the chiral fermion wavefunctions amounts to
an anomaly-cancelling term in the
path integral of chiral gauge theory~\cite{2}.  That the theory is not
of the conventional form can also be
seen in the interaction picture~\cite{3}.  In this picture, the time
displacement operator $U(t,0)$
is subjected to the Schwinger-Tomonaga equation whose solution is routinely
given as the time-ordered product
\begin{eqnarray}
U(t,0) &=& {\cal T}\exp\left\{ -i\int^t_0 H_{\rm int}(\tau)d\tau\right\},
\label{1}
\end{eqnarray}
where $H_{\rm int}$ is the interacting part of the Hamiltonian.  From this
expression, statements of anomalous symmetry breaking of chiral gauge theory
necessarily follow.  However, expression~(\ref{1}) is not always the true
solution.  A counter-example is given in reference~\cite{4}.  Chiral gauge
theory is another important exception.

Explicit calculation has in fact picked up an extra term for the exponential
argument of~(\ref{1}).  In two dimensions, the term is of the form
\begin{eqnarray}
\int dx (A_0 + A_1)\frac{\partial_1}{\partial_0 + \partial_1}(A_0+A_1),
\label{1a}
\end{eqnarray}
for abelian theory in which the gauge fields couple, with coupling constant
$e$,
only to the left-handed current, projected out by $P_L = \frac{1}{2}
(1-\gamma_5)$, $\gamma_5 = \gamma_5^\dagger$.
The term~(\ref{1a}) generates under the
gauge transformations the anomaly cancellation and so preserves the gauge
invariance of the theory: chiral gauge theory is anomaly-free by itself
without the need of outside arrangement.  It is emphasised that these terms
are not introduced by hands in an {\it ad hoc} manner but are the result
of a careful treatment of the theory.  Their non-locality places them
outside the class of allowed local counterterms.
Being anomaly-free, the gauge current of the theory is now conserved.  It is
the {\it total} dynamical current,
consisting of the chiral fermion current {\it and} another contribution due
to the coupling in~(\ref{1a}).  The
chiral fermion current is still not conserved since the presence of~(\ref{1a})
cannot modify the Feynman rules involving fermions and, in particular, the
results of the well-known anomalous fermion loops and of the anomalies, if any,
of global symmetry.  Nevertheless, the theory is consistent
as will be argued for shortly.

Expression~(\ref{1a}) contains an anomaly-cancelling, Wess-Zumino part
\[\left[\frac{\partial_\mu A^\mu}{\Box}\right]\epsilon_{\rho\sigma}
\partial^\rho A^\sigma.\]
The Green's function $\frac{1}{\Box}$ is uniquely defined with
appropriate boundary condition; however, owing to the reality of the action
it cannot contain an imaginary part so the principal value prescription
will be adopted here.  There is another ambiguity for $\frac{\partial A}
{\Box}$ under the gauge transformation with the parameter $\chi$ satisfying
\begin{eqnarray}
\Box \chi &=& 0,
\label{2a}
\end{eqnarray}
because of the ambiguity of the product of distributions, in momentum space,
$ \frac{1}{p^2}p^2\delta(p^2) = ?$.
To fix the ambiguity one has to conform with the transformation of~(\ref{1a}),
and requires
\begin{eqnarray}
\left[\frac{\partial A}{\Box}\right] &\to& \frac{1}{e}\chi.
\label{2b}
\end{eqnarray}
The definition of the square bracket notation can be imposed consistently
in deriving the Feynman rules from the action and will be assumed
implicitly from now on.

We introduce the change of variables
\begin{eqnarray}
\psi'(x) &=& (h(x)P_L + P_R)\psi(x),\nonumber\\
\bar\psi'(x) &=& \bar\psi(x)(P_L + h^\dagger(x) P_R),
\label{3}
\end{eqnarray}
with
\begin{eqnarray}
h(x) &=& \exp\left\{ -ie\frac{\partial A(x)}{\Box}\right\},
\label{33}
\end{eqnarray}
upon which the new fermionic fields are invariant with respect to the
gauge transformations
\begin{eqnarray}
\psi(x) &\to& (g(x)P_L +P_R)\psi(x),\nonumber\\
\bar	\psi(x) &\to& \bar\psi(x)(P_L + g^\dagger(x) P_R),\\
A_\mu(x) &\to& g(x)A_\mu(x)g^\dagger(x) - i\frac{1}{e}\partial_\mu g(x)
g^\dagger(x).
\nonumber
\label{4}
\end{eqnarray}
The change of variables~(\ref{3}) itself is a gauge transformation and
associated with it is the Fujikawa jacobean~\cite{fuji} from the fermionic
measure of the path integral.  The term~(\ref{1a}) is then cancelled by
the jacobean, up to local, regulator-dependent counterterms.  The resulted path
integral is
\begin{eqnarray}
{\cal Z} &=& \int [d\bar\psi'][d\psi'][d A_\mu]\exp\{ i{\cal S}\},
\label{4a}
\end{eqnarray}
with the action
\begin{eqnarray}
{\cal S}&=&\int dx\left( -\frac{1}{4}F^2 + \bar\psi'(i\dslash +
e\aslash' P_L)\psi' + \frac{m^2}{2}A'_\mu A'^{\mu} + M\bar\psi'\psi' \right),
\label{5}\\
A'_\mu(x) &:=& h(x)A_\mu(x)h(x)^\dagger - i\frac{1}{e}\partial_\mu h(x).
\nonumber
\end{eqnarray}
For abelian gauge group, we have
\begin{eqnarray}
A'_\mu &=& \left(g_{\mu\nu}-\frac{\partial_\mu \partial_\nu}{\Box}
\right) A^\nu.
\label{55}
\end{eqnarray}
With the appearance of non-local operator on $A_\mu$,
non-local counterterms will be generated.  However, they are severely
constrained
by the gauge symmetry.  For that reason a gauge-invariant mass term for the
transverse components has been included in the action above, as has a fermion
mass term.

The theory~(\ref{5}) is explicitly gauge invariant with
the transformations of the bosonic gauge fields, as such transformations
leave the functional measure unchanged.  And that is all that counts in a
gauge theory in order to eliminate the unphysical degrees of freedom of the
gauge fields.  Even though the discussion so far is in two dimensions, we also
propose the theory for abelian, four-dimensional
chiral gauge theory.  The four-dimensional Wess-Zumino has the form
\[ \frac{\partial A}{\Box}F\tilde F \]
and is cancelled out in the path integral~(\ref{4a}) and~(\ref{5}).

For non-abelian gauge group, one can go back to the Schwinger-Tomonaga equation
to find the modification to the time-ordered time displacement operator.
Alternatively, it suffices
to find the function $h(x)$ which takes the value in the gauge
group and leaves the fermionic fields in~(\ref{3}) gauge-invariant.
Namely, we want $h(x)$ gauge transforms so that
\begin{eqnarray}
h^\dagger(x)\delta h(x) = -iT^a \epsilon^a(x),
\label{555}
\end{eqnarray}
with the infinitesimal gauge parameter $\epsilon$ and the convention
\begin{eqnarray}
[T^a,T^b] &=& if^{abc}T^c,\nonumber\\
{\rm tr}\{T^aT^b\} &=& -\frac{1}{2}\delta^{ab}.\nonumber
\label{7a}
\end{eqnarray}
Condition~(\ref{555}) only defines $h(x)$ up to an ambiguous gauge-invariant
part.  To remove the ambiguity, we demand that $h(x)$ reduces to~(\ref{33})
for abelian gauge group and thus
impose the constraint that $h(x) = 1$ in the gauge
\begin{eqnarray}
\frac{\partial A^a}{\Box} = 0.
\label{10}
\end{eqnarray}
This constraint completely fixes the gauge as can be seen
from~(\ref{2b}), whereas in the Lorentz gauge $\partial A = 0$ some residual
transformations of the type~(\ref{2a}) are still permitted.

Such a non-abelian $h(x)$ can be evaluated order by order in the
coupling constant to be~\cite{5}
\begin{eqnarray}
h(x) &=& \exp\left\{ -ie\frac{\partial A}{\Box} + ie^2T^af^{abc}\frac{1}{\Box}
\left[ \partial_\mu\left(A_\mu^b\frac{\partial A^c}{\Box} \right)
-\frac{1}{2}\partial A^b\frac{\partial A^c}{\Box}\right] \right.\nonumber\\
&& + \left.O(e^3) \right\},
\label{8}
\end{eqnarray}
This non-abelian change of variables once again results in~(\ref{5}) but
with trivial notation for non-abelian theory and with
\begin{eqnarray}
A'^a_\mu &=& \left(g^{\mu\nu}-\frac{\partial^\mu\partial^\nu}{\Box} \right)
\left\{A^a_\nu + ef^{abc}\left[\frac{1}{2}\left(\partial_\nu\frac{\partial A^b}
{\Box}\right) - A^b_\nu\right]\frac{\partial A^c}{\Box}
\right.\nonumber\\
&& \left.+ O(e^2)\right\}.
\label{9}
\end{eqnarray}
In particular, the gauge-fermion coupling in~(\ref{5}) now has the explicit
form
\begin{eqnarray}
eA_\mu^{'a}\bar\psi'T^a\gamma^\mu P_L\psi',
\label{7}
\end{eqnarray}
and is gauge invariant up to $O(\epsilon e^2)$, $\epsilon$ being the gauge
transformation parameter, as $A'_\mu$ given by~(\ref{9}) is invariant
to that order.
Attention is drawn to the fact that higher order
corrections to $A'_\mu$ is always subjected to the tranversality projector.

In undoing the change of variables~(\ref{3}) the non-abelian
Wess-Zumino term can be obtained through the evaluation of the Fujikawa
jacobean.  Its precise form is complicated, particularly for semi-simple
group like $SU(2)\times U(1)$, but is not required for our
arguments.
The perturbative definitions~(\ref{8}) and~(\ref{9}) above take on simple
forms in the gauge~(\ref{10}), whence
$A'_\mu = A_\mu$, $\psi' = \psi$, because the left hand side
of~(\ref{10}) always appears in higher corrections to~(\ref{8}) and~(\ref{9}).
The fact that in this gauge the theory~(\ref{5}) is manifestly
local, up to the Faddeev-Popov determinant which can be localised by the
introduction of ghosts, should convince us that our theory, despite of
its appearance, is not more non-local than conventional quantum field theory.

The theory~(\ref{5}) with appropriate fields in the abelian case~(\ref{55})
and non-abelian~(\ref{9}) is our proposal for four-dimensional chiral gauge
theory.  The transverse mass term in~(\ref{5}) is
consistent for abelian gauge group, where the transverse and longitudinal
components separate, if the current is conserved as will be argued for below.
The non-abelian case is more complicated since the two components of the
gauge fields no longer decouple and consequently the integration over the
longitudinal components is not factorised in the path integral.
However, the non-abelian vector boson transverse mass term is
gauge invariant and thus is not subjected to the
criticism of non-renormalisability of non-invariant massive Yang-Mills
theory~\cite{IZ}.  In fact, the vanishing mass limit can be taken
smoothly in ~(\ref{5}) unlike in the latter case.

The theory is formally gauge invariant and should be renormalisable and
unitary.  The only danger for the generalised Ward identities comes from the
fermion triangle loop, which is not renormalised by higher order corrections
owing to the non-renormalisation theorem for chiral anomalies~\cite{6}.
The chiral fermion currents are not covariantly conserved as can be seen from
the triangle diagrams~\cite{7}
\begin{eqnarray}
k_{(3)}^\rho\Gamma_{\mu\nu\rho}^{abc}(k_{(1)},k_{(2)},k_{(3)}) &=&
-d^{abc}\frac{e^3}{12\pi^2}\epsilon_{\mu\nu\alpha\beta}k_{(1)}^\alpha
k_{(2)}^\beta.
\label{12a}
\end{eqnarray}
It is customarily demanded that $d^{abc} := {\rm tr}\{T^a,T^b\}T^c = 0$ or
$\sum e^3 = 0$.
Such a requirement, however, is not necessary here.  It is seen from~(\ref{55})
and~(\ref{9}) that the transversality projector
$\left[g_{\mu\nu}-\frac{k_\mu k_\nu}{k^2}\right]$
can be transferred onto the fermion currents in the gauge-fermion
coupling~(\ref{7}), modifying the interaction vertex~\cite{tdk}.
The anomalous triangle loop then decouples when the external boson is
longitudinal since the loop is
flanked at each of its three vertices with the transversality projector.
The other potentially dangerous diagrams beside the triangle
can be shown~\cite{8} to be non-anomalous once the anomalous triangle
decouples.  The additional contribution of non-zero fermion mass to the
loops should not spoil the generalised Ward identities involving the gauge
fields.


We thus argue that the generalised Ward identities, derived with
appropriate gauge fixing and source terms, are maintained throughout
up to terms vanishing together with the employed regulator.  The lack of a
chiral gauge invariant regulator is no worse than the situation in the
existing proof
of renormalisability and unitarity of the Standard Model.  Like that
proof, the annihilation of the triangle diagram, annihilated automatically for
the case presented herein, is sufficient.  But unlike the usual situation,
we have non-local terms.  However, the non-locality cannot
render the theory non-renormalisable owing to the constraint of gauge
invariance.

The well-known bad high-energy behaviour of the Born scattering amplitudes for
non-invariant massive Yang-Mills~\cite{Taylor} is ultimately
connected to the non-renormalisability of that theory.
And the equivalence theorem for
massive Yang-Mills~\cite{LLCorn} demands the existence of the Higgs sector.
While there is no apparent scalar coupling in~(\ref{5}), there are couplings
in the original fields of the forms
\begin{eqnarray}
-ieM\left(\frac{\partial A^a}{\Box}\right)\bar\psi T^a\gamma_5\psi
\end{eqnarray}
and
\begin{eqnarray}
-emf^{abc}\left[\left(g^{\mu\nu}-\frac{\partial^\mu\partial^\nu}
{\Box}\right)A_\mu^a\right]\left[\left(\frac{1}{2}\left(\partial_\nu\frac{\partial A^b}
{\Box}\right)-A_\nu^b\right)\frac{\partial A^c}{\Box}\right],
\label{13}
\end{eqnarray}
in which $\frac{\partial A^a}{\Box}$ simulates~\cite{Vtrick} the effects of
scalar fields, with coupling strength proportional to the various masses,
to absorb the longitudinal components as demanded by the theorem.  Note that
we have not introduced and new dynamical field in constrast to the case of
radially-fixed Higgs, which is untenable because the radial components
will inevitably be revived non-perturbatively.

It remains to be seen whether the
approach presented is realised in the electroweak interactions.

I am indebted to Herbert Neuberger for many illuminating discussions.
I also wish to thank Steve Adler, Freeman Dyson, Samson Shatashvili and
Frank Wilczek for discussions; the IAS for the support and hospitality
during my stay.  This work is partially supported by the Australian Research
Council.


\begin{thebibliography}{99}
\bibitem{Vtrick} Their effective propagators are obtained from those of
the massive vector bosons with the help of the Veltman trick
\[ (k^\mu/k^2)[(g_{\mu\nu}-k_\mu k_\nu/m^2)/(k^2-m^2+i0^+)](k^\nu/k^2)
=-1/(k^2 m^2).\]
Also note that the principal value prescribed to $(1/\Box)$ being real
cannot contribute as a physical particle of mass zero to the S matrix.
\end{thebibliography}
\end{document}